\documentclass{nature-jae}
\usepackage{graphics,psfig,epsfig}

\def\la{\mathrel{\hbox{\rlap{\hbox{\lower4pt\hbox{$\sim$}}}\hbox{$<$}}}}
\def\ga{\mathrel{\hbox{\rlap{\hbox{\lower4pt\hbox{$\sim$}}}\hbox{$>$}}}}

\title{Water Vapor and Hydrogen in the Terrestrial Planet Forming
Region of a Protoplanetary Disk}

\author{J.A. Eisner}

\begin{document} 

\maketitle
\begin{affiliations}
\item Department of Astronomy, 601 Campbell Hall,
University of California, Berkeley, CA 94720
\end{affiliations}

\begin{abstract}
Planetary systems, ours included, are formed in disks of dust and
gas around young stars.  Disks are an integral part of the star 
and planet formation process\cite{SAL87,SAFRONOV69}, and knowledge of the 
distribution and temperature of inner disk material is 
crucial for understanding terrestrial
planet formation\cite{RQL04}, giant planet migration\cite{LBR96},
and accretion onto the central star\cite{KONIGL91}.  
While the inner regions of protoplanetary disks in nearby star forming regions 
subtend only a few nano-radians, near-IR interferometry
has recently enabled the spatial resolution of these terrestrial
zones.  Most observations have probed only dust\cite{MILLAN-GABET+07}, 
which typically dominates the near-IR emission.
Here I report spectrally dispersed near-IR interferometric observations
that probe gas (which dominates the mass and dynamics of the inner disk), 
in addition to dust, within one astronomical unit of the young star MWC 480.  
I resolve gas, including water vapor and atomic hydrogen, interior to the edge 
of the dust disk; this contrasts with results of previous spectrally dispersed 
interferometry observations\cite{MALBET+07,TATULLI+07}.
Interactions of this accreting gas with migrating planets may lead to 
short-period exoplanets like those detected around main-sequence
stars\cite{LBR96}.  The observed water vapor is likely produced by the 
sublimation of migrating icy bodies\cite{CC06}, and provides a potential 
reservoir of water for terrestrial planets\cite{DRAKE05}.
\end{abstract}

I obtained observations with a new grism\cite{EISNER+07b}  
(a dispersive element consisting of a diffraction grating 
on the vertex of a prism) 
at the Keck Interferometer\cite{CW03,COLAVITA+03}
that show, for the first time, evidence of spatially resolved gaseous 
spectral-line emission interior to the edge of the dusty disk around a young 
star, MWC 480.  My observations provide a
spectral dispersion of $R=230$ from 2.0 to 2.4 $\mu$m wavelength,  
and an angular resolution of approximately 1 milli-arcsecond 
(mas; 1 mas $\approx$ 5 nano-radians), which is $\sim 0.1$ astronomical units
(AU) at the 140 parsec distance to the target (the source is located in the
Taurus-Auriga star forming complex).
MWC 480 has a stellar mass of $2.3$ M$_{\odot}$, an age of 
$\sim 6$ Myr, and a circumstellar 
disk of dust and gas in Keplerian rotation\cite{MKS97}. Previous 
observations found that the dusty component of the disk extended to within 
$\sim 0.5$ AU of the central star\cite{EISNER+04,MONNIER+06}, and suggested 
the presence of hot gaseous emission at even smaller stellocentric 
radii\cite{EISNER+07a}.  My observations confirm these suggestions, and
resolve hydrogen gas and water vapor in the inner disk.

My interferometry data provide a measurement of the angular size of the 
circumstellar emission from MWC 480 as a function of wavelength (Figure 1). 
The spectral dependence of the angular size indicates a radial temperature 
gradient.  Since cooler material emits relatively more radiation
at longer wavelengths,  an outwardly decreasing temperature gradient yields
an increasing fraction of emission from larger radii, and hence a larger
angular size, at longer wavelengths.
The temperature gradient probably arises from multiple components at different 
stellocentric radii\cite{EISNER+07a}: warm dust at radii where temperatures
approach the dust sublimation temperature, and hotter gas at smaller radii.

Figure 1 also shows more compact emission at 2.165 $\mu$m relative
to the surrounding continuum.  This dip corresponds to the wavelength of the
Br$\gamma$ electronic transition ($n=7 \rightarrow 4$) of hydrogen.
The Br$\gamma$ emission observed from MWC 480 traces very hot 
gas ($\sim 10^4$ K) that has recently recombined from 
ionized hydrogen\cite{OSTERBROCK89}, presumably in the inner 
reaches of an accretion column.


To match the features of Figure 1, I constructed a simple physical model and 
fitted it directly to the measured $V^2$ and fluxes.  The model
accounts for emission of the unresolved central star, circumstellar 
continuum emission from hot gas, spectral-line emission from hydrogen 
(Br$\gamma$), and continuum emission from warm dust.  
I take each emission component to be a uniform ring whose 
width is 20\% of its radius.  This multiple-ring model approximates a 
continuous disk where dust, gas, and individual gaseous atoms or molecules 
contribute emission at different size scales. It has been shown previously 
that ring models provide reasonable approximations to more complex 
models of inner disk regions\cite{EISNER+04,IN05}.
I also assume that the gaseous continuum and Br$\gamma$ emission originate
from stellocentric radii smaller than 0.1 AU; given the $\sim 1$ mas angular 
resolution of my observations, I can not accurately distinguish 
between smaller radii, although I can rule out larger radii.  

As indicated in Figure 2,
the model fits the data well over most of the spectral bandpass.  
However, there are substantial deviations between the model predictions
and the data in the region between 2.0 and 2.1 $\mu$m (particularly for the
fluxes).  These deviations suggest the presence of hot water vapor, which has 
substantial opacity from 2.0 to 2.1 $\mu$m, in the inner disk 
of MWC 480.  I include water vapor in the model by adding a ring of
material with the opacity of water vapor\cite{LUDWIG71}.  The superior quality 
of the fit when water is included clearly demonstrates the presence of H$_2$O 
in the inner disk of MWC 480 (for further arguments why the model
must include water vapor emission in order to match the data, see the 
Supplementary Information section).

The best-fit model includes
compact gaseous continuum emission with a temperature of 
$2410 \pm 125$ K and Br$\gamma$ emission
that adds $17\% \pm 7\%$ to this gaseous continuum flux at 2.165 $\mu$m; these 
components are situated at stellocentric radii $<0.1$ AU.  Water vapor
emission is located at a radius of $0.16 \pm 0.05$ AU.  The temperature of 
this material is $2300 \pm 120$ K and the column density is 
$1.2 \pm 0.6 \times 10^{19}$ cm$^{-2}$.  Dust emission 
originates from a radius of $0.28 \pm 0.01$ AU and has a temperature of 
$1200 \pm 60$ K.  Quoted 1$\sigma$
uncertainties are quadrature sums of statistical 
uncertainties in the fits and assumed 5\% systematic errors.


I illustrate the basic geometry of the innermost regions of MWC 480 in 
Figure 3.  As inferred from the data, gaseous emission is found at 
substantially smaller stellocentric radii than the dust. The inner edge of the 
dust disk lies at $\sim 0.3$ AU, where the disk temperature approaches the
sublimation temperature of silicate dust\cite{POLLACK+94}.
Because I also include contributions from gas in my analysis, the inferred
dust size is larger, and the temperature lower, than in previous work where
only dust emission was considered\cite{EISNER+04,MONNIER+06}.
Emission from H$_2$O is
found at $\sim 0.15$ AU, and hot gaseous continuum and Br$\gamma$ emission
at still smaller radii ($<0.1$ AU).  
The observed Br$\gamma$ emission probably originates in an accretion column, 
near the shock formed when infalling material
impacts high latitude regions of the star\cite{KONIGL91,NCT96,CG98}.  

My results substantiate models of inner disk structure 
proposed previously for similar star+disk systems\cite{MUZEROLLE+04}.  
Comparison with these models suggests that
in order to reproduce the ratio of flux in the gas and dust components
determined here ($\sim 0.5$), a mass accretion rate $\ga 10^{-7}$
M$_{\odot}$ yr$^{-1}$ must be invoked for MWC 480.  At such high accretion
rates, the inner disk gas becomes optically thick to its own radiation, and
can therefore produce relatively more flux.  Furthermore, models with this
high accretion rate produce strong Br$\gamma$ emission, consistent with my 
detection.  My measurement of gas interior to the
edge of the dust disk also confirms previous suggestions of such structure
based on modeling of spatially unresolved (but spectrally resolved) 
observations of gaseous emission lines\cite{NCT96,CTN04,NAJITA+06} or very 
low dispersion interferometric measurements\cite{EISNER+07a}.

The finding that gaseous emission, and Br$\gamma$ emission in particular,
lies within the edge of the dusty component in protoplanetary disks contrasts
with previous interferometric studies of young stars.  
Previous work found the Br$\gamma$ emission to
lie at radii equal to or larger than the dust, implying an origin of the 
emission in an outflowing wind\cite{MALBET+07,TATULLI+07}.
The differing results
for MWC 480 and the two previously observed sources may stem from
different luminosities of the objects.  MWC 480 is orders of magnitude
less luminous than MWC 297\cite{EISNER+04}, for which the Br$\gamma$ emission
is more extended than the dust emission\cite{MALBET+07}, and approximately half
as luminous at HD 104237\cite{VDA+98}, for which the Br$\gamma$ and dust 
emission have comparable size scales\cite{TATULLI+07}.  Br$\gamma$ emission
probably traces both accreting and outflowing components in all of these
sources to some extent, but the accretion appears to increasingly dominate the 
emission for lower-luminosity stars.

The distribution of gas at stellocentric radii $<0.1$ AU in the inner disk of 
MWC 480 shows that material is present in this protoplanetary disk at radii 
comparable to the semi-major axes of short-period exoplanets\cite{MARCY+05}.  
This finding supports the
hypothesis that giant planet migration is halted in resonances with the 
inner edge of the gaseous disk\cite{LBR96,NAJITA+06}.  Future observations of
gaseous disk truncation radii in solar-type T Tauri stars will enable
direct comparison with exoplanets, which are predominantly found around 
solar-type stars.  

The presence of water vapor in the inner regions of the 
protoplanetary disk surrounding MWC 480 constrains how water 
(and other material) is transported in the disk terrestrial zone. 
Because water beyond the ice-line is in solid form, a concentration gradient
tends to push water vapor out of the inner disk; the fact that I observe 
this inner disk vapor suggests continual replenishment through the sublimation
of inwardly migrating icy bodies\cite{CC06}.  Water vapor in disk
terrestrial regions may also enable the {\it in situ} formation of 
water-rich planets via adsorption of water onto dust grains and subsequent
growth of these water-laden grains into planetesimals\cite{DRAKE05}.
Finally, as one of the dominant gaseous opacity sources in the near-IR,
water vapor can provide an excellent tracer of the temperature and
velocity structure of inner disk gas\cite{CTN04}.


\bibliography{jae_ref}

\begin{addendum}
 \item Data presented herein were obtained at the W.M. Keck Observatory from 
telescope time allocated to the National Aeronautics and Space Administration 
through the agency's scientific partnership with the California Institute of 
Technology and the University of California. The Observatory was made possible 
by the generous financial support of the W.M. Keck Foundation.  I thank the
entire Keck Interferometer team for their invaluable contributions to these
observations. I also acknowledge input into this work (and this manuscript)
from R. Akeson, E. Chiang, A. Glassgold, J. Graham, J. Najita, and R. White.
The author is supported by a Miller Research Fellowship.
 \item[Competing Interests] The author declares that he has no
competing financial interests.
 \item[Correspondence] Correspondence and requests for materials
should be addressed to J.A.E. \\ (email: jae@astro.berkeley.edu).
\end{addendum}

\clearpage

\includegraphics[clip=true,scale=1.0]{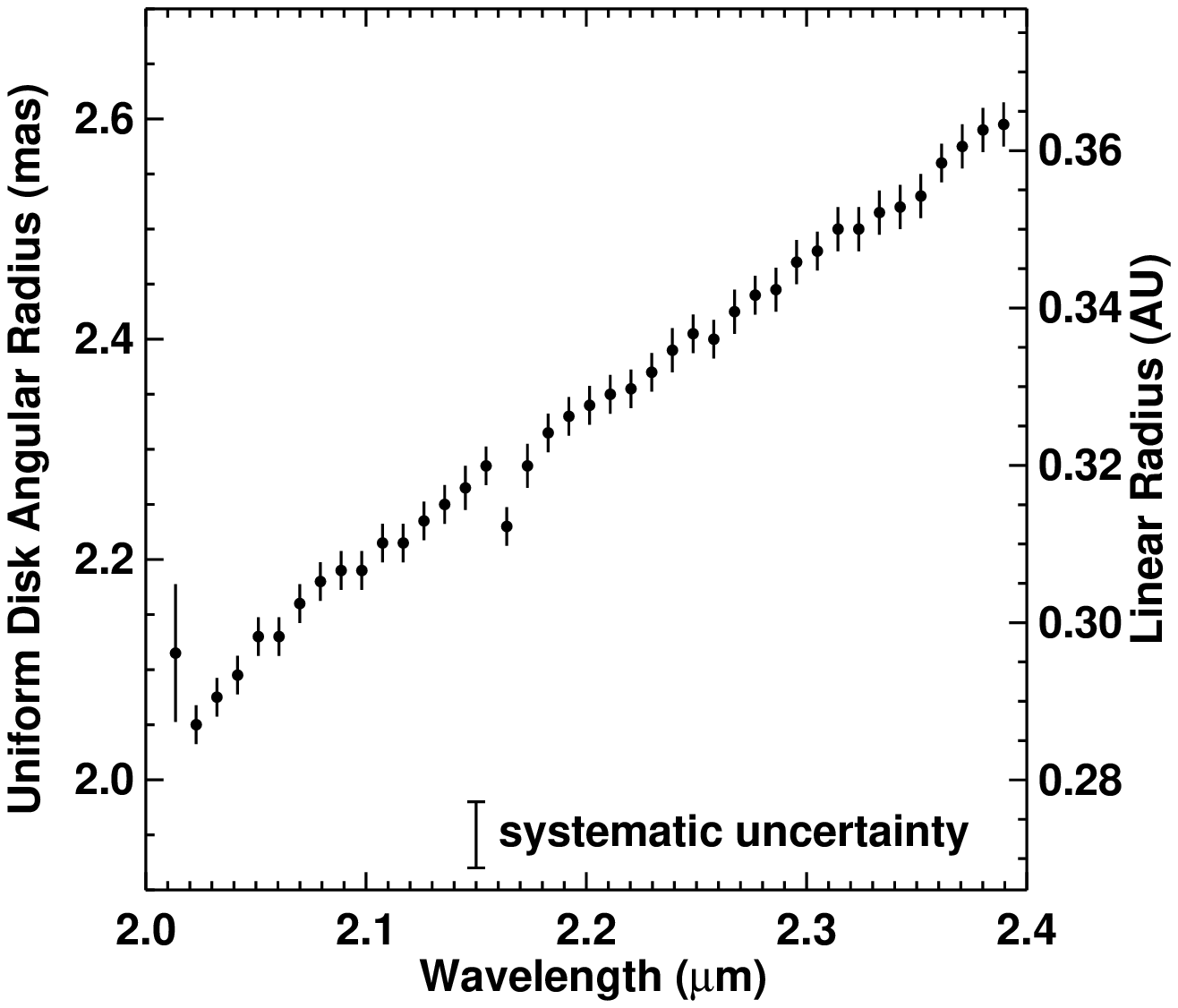}
\newline
\noindent {\bf Fig. 1.}
{\bf Angular size of the near-IR emission from MWC 480 as a function of
wavelength.}   Sizes are calculated from squared visibility measurements
($V^2$) made with the Keck Interferometer grism on 12 Nov 2006.   
The $V^2$ (and fluxes; see Fig. 2), were calibrated using 
observations of unresolved stars of known spectral type and 
parallax\cite{EISNER+07b}.  I applied these calibrations to unresolved 
``check stars'' to test the calibration, and thereby estimated 
channel-to-channel uncertainties ($1\sigma$ standard deviation) in my flux and
$V^2$ measurements of 3\%.  
The shortest-wavelength data point has a larger uncertainty, because 
atmospheric water vapor absorption
leads to lower photon counts in this channel; I estimate the uncertainty
in this channel at $\sim 10\%$.
There is also a systematic uncertainty of approximately 5\% in the overall 
normalization of the data, due to errors in 
the absolute calibration\cite{EISNER+07b}.
I assume that the emission morphology is a uniform disk, 
for which the measured $V^2$ are
related to the angular diameter of the source, $\theta_{\rm UD}$, by
$V^2 = \left\{{J_1(\pi \theta_{\rm UD} B_{\rm proj}/\lambda)}
/({\pi \theta_{\rm UD} B_{\rm proj}/\lambda})\right\}^2,$
where $B_{\rm proj}$ is the projected baseline separation of the two telescopes
as seen by the source, $\lambda$ is the wavelength, and $J_1$ is a first-order
Bessel function.  I remove the stellar contribution to the 
visibilities\cite{EISNER+04}, and the plotted sizes thus represent only
the circumstellar emission.  The angular size increases from
2.0 to 2.4 $\mu$m, indicating multiple emission components: hot gas
at small stellocentric radii and warm dust at larger radii\cite{EISNER+07a}.  
In addition, the angular size 
is smaller at 2.165 $\mu$m, the wavelength where hot hydrogen gas emits in the 
Br$\gamma$ line,  than at adjacent continuum wavelengths.  

\includegraphics[clip=true,scale=1.0]{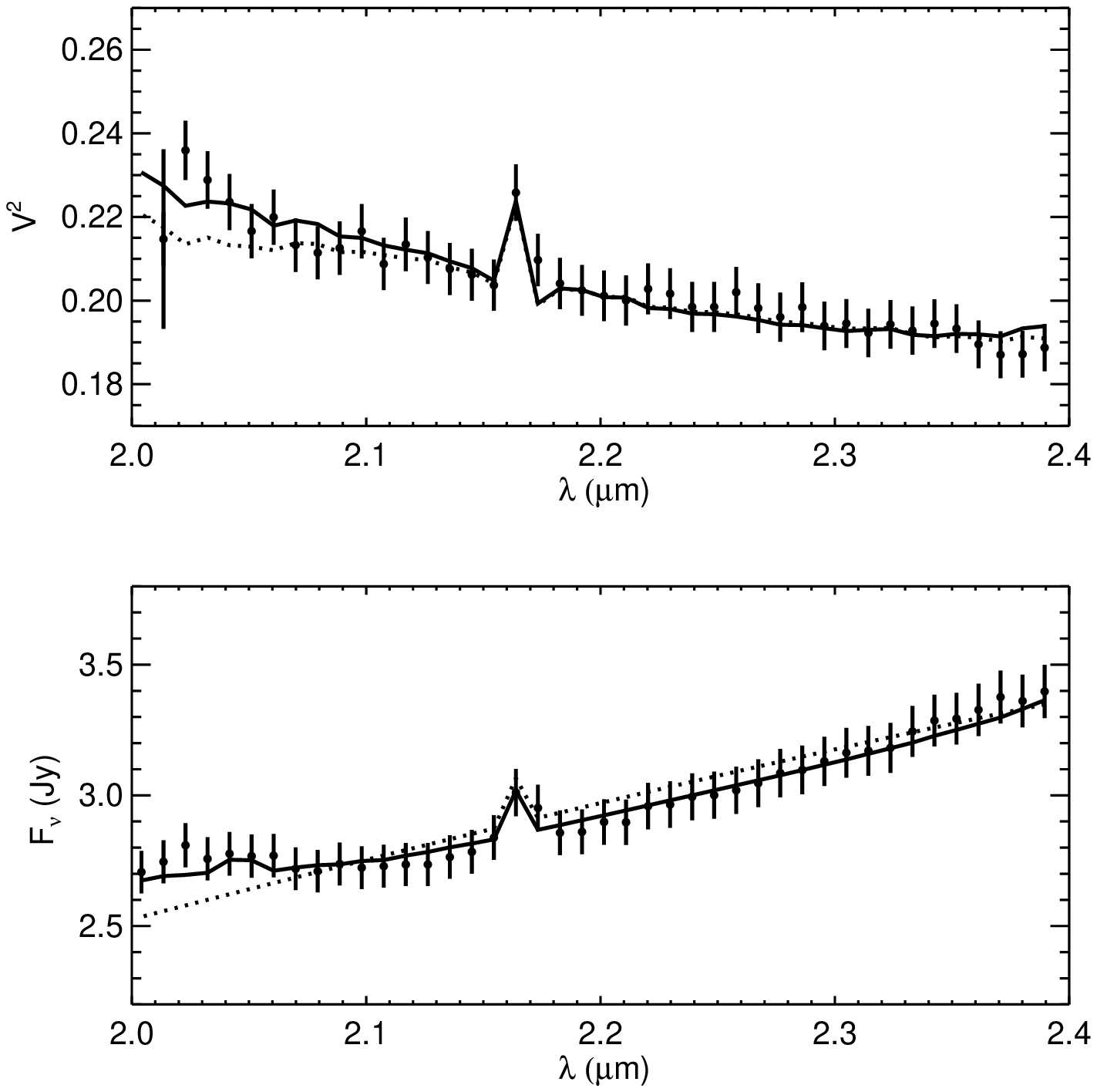}
\newline
\noindent{\bf Fig. 2.}
{\bf Measured $V^2$ and fluxes, compared to the 
predictions of simple physical models.}  Plotted error bars indicate
estimated 1$\sigma$ (standard deviation) uncertainties of 3\%; as noted
above, the uncertainty in the shortest wavelength $V^2$ measurement is
larger, approximately 10\%.  
The dotted line shows the values computed for a model including continuum
emission from warm dust and hot gas, and Br$\gamma$ emission from hydrogen. 
This model reproduces the overall slopes seen in the data as well as the 
feature at 2.165 $\mu$m, but deviates from the data between 2.0 and 2.1
$\mu$m.  Hot water vapor contributes significant emission at these 
wavelengths\cite{LUDWIG71}, and the solid line shows the predictions of a 
model that includes H$_2$O.  Because the H$_2$O emission lies at a radius only
slightly larger than that of the hot gaseous continuum, its effect on the 
visibilities is more subtle than its effect on the measured fluxes. 
Absorption due to terrestrial atmospheric
water vapor affects the data in the shortest-wavelength channel, but not
substantially in longer-wavelength channels since, in contrast to the 
hot water vapor in our model, water vapor at atmospheric temperature and
pressure conditions lacks substantial opacity at these 
wavelengths\cite{LUDWIG71}. The model including H$_2$O fits the data well 
over the entire observed wavelength range, and yields substantially smaller
residuals between model and data than the model without water vapor.
We note that the inferred sizes of the emission components in our model
are smaller than the uniform disk sizes plotted in Figure 1; this is
due primarily to the different characteristic sizes for uniform disk and
relatively narrower ring models\cite{EISNER+04}.
Additional details of the modeling and fitting
procedure can be found in previous work\cite{EISNER+07a}.  

\clearpage
\includegraphics[clip=true,scale=1.0]{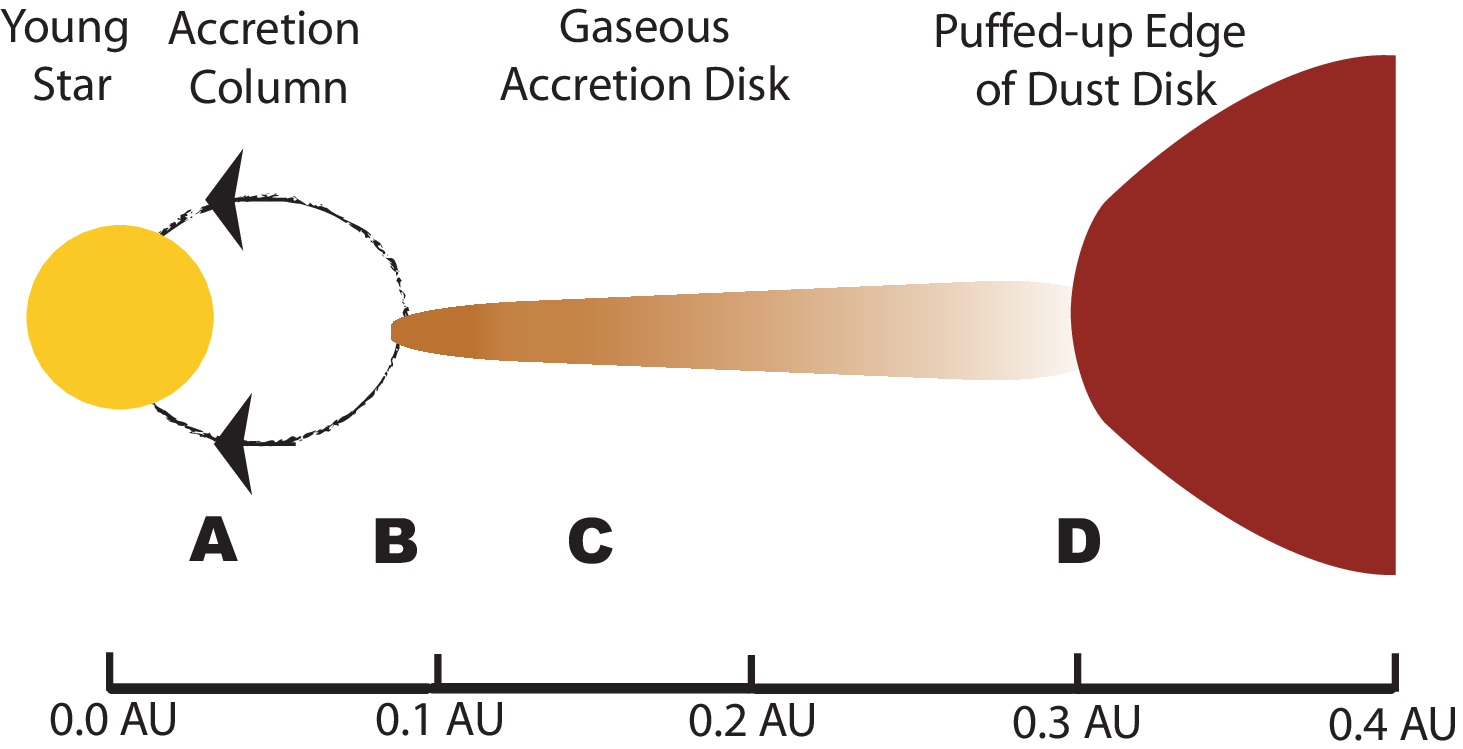}
\newline
\noindent {\bf Fig. 3.} 
{\bf The environment within 1 AU of the young star MWC 480.}
Dust in the inner disk around MWC 480 extends inward to  
approximately 0.3 AU ({\bf D}), where the disk temperature becomes
high enough to sublimate the dust.  The ring-like appearance of the 
dust sublimation front is compatible with my data, and with physical models 
predicting a puffed-up and curved inner edge\cite{DDN01,IN05}.  Interior to 
the dust sublimation radius, gaseous material continues to accrete onto the 
young star (and may also be blown away in a tenuous 
wind\cite{SHU+94}, which is not depicted).  
Based on a sketch for a similar young star\cite{MUZEROLLE+04}, 
I have portrayed a continuous gaseous disk whose optical depth
increases at smaller radii.  I observe hot continuum
emission from the hottest, densest gaseous material at stellocentric radii 
smaller than 0.1 AU
({\bf B}) and spectral-line emission from hot water vapor at slightly larger
radii ({\bf C}).  Finally, I observe Br$\gamma$ emission at radii smaller 
than 0.1 AU, which traces hydrogen gas at high temperatures 
($\sim 10^4$ K).  I speculate that 
this emission arises in the hot accretion shock where infalling gas impacts 
the stellar surface ({\bf A}).  The $\sim 1$ mas angular resolution of
my observations translates into a spatial resolution of approximately 0.1 AU
at the distance to MWC 480.

\clearpage





\end{document}